\documentclass[a4paper,11pt,dvips]{article}
\usepackage{graphicx}
\usepackage{amsmath}
\usepackage{amssymb}
\usepackage{latexsym}

\newcommand{\beq}{\begin{equation}}
\newcommand{\eeq}{\end{equation}}
\newcommand{\beqn}{\begin{eqnarray}}
\newcommand{\eeqn}{\end{eqnarray}}

\newcommand{\te}{\theta}

\newcommand{\pa}{\partial}

\newcommand{\del}{\delta}

\newcommand{\da}{\dagger}

\newcommand{\lb}{\label}

\newcommand{\NP}[1]{ {\it Nucl.~Phys.} {\bf #1}}

\newcommand{\PR}[1]{ {\it Phys.~Rev.} {\bf #1}}
\newcommand{\PRL}[1]{ {\it Phys.~Rev.~Lett.} {\bf #1}}

\newcommand{\IJMP}[1]{ {\it Int.~J.~Mod.~Phys.} {\bf #1}}
\newcommand{\JP}[1]{ {\it J.~Phys.} {\bf #1}:\  Math.~Gen.~}

\begin{document}
\begin{titlepage}
\setcounter{page}{1}
\renewcommand{\thefootnote}{\fnsymbol{footnote}}

\begin{flushright}
UCDTPG 08-01\\
arXiv:0805.2388
\end{flushright}

\vspace{5mm}
\begin{center}

{\Large\bf Fractional Quantum Hall States in Graphene}

\vspace{5mm}

{\bf Ahmed Jellal$^{a,b,}$\footnote{ajellal@ictp.it, jellal.a@ucd.ac.ma}}
and {\bf Bellati Malika$^{b,}$\footnote{mbelati@gmail.com}}

\vspace{5mm}

{\em $^{a}$The Abdus Salam International Centre for Theoretical Physics},\\
{\em  Strada Costiera 11, 34014 Trieste, Italy}\\


{ \em $^{b}$Theoretical Physics Group,  
Faculty of Sciences, Choua\"ib Doukkali University},\\
{\em Ibn Ma\^achou Road, PO Box 20, 24000 El Jadida,
Morocco}\\

\vspace{30mm}

\begin{abstract}

We quantum mechanically analyze the  fractional quantum Hall
effect in graphene. This will be done by
building the corresponding  states in terms of
a potential governing the interactions and discussing other issues.
More precisely, we consider a system of particles
in the presence of an external
magnetic field and take into account of a specific
interaction that captures the basic features of the
Laughlin series $\nu={1\over 2l+1}$. We show that how its
Laughlin potential can be generalized to deal with
the composite fermions in graphene. To give a concrete
example, we consider the $SU(N)$ wavefunctions and
give a realization of the composite fermion
filling factor. All these results will be obtained by generalizing
the mapping between the Pauli--Schr\"odinger and Dirac Hamiltonian's
to the interacting particle case. Meantime by making use of a gauge transformation,
we establish a relation between the free and interacting Dirac operators.
This shows that the involved interaction can actually be generated from a
singular gauge transformation.

\end{abstract}
\end{center}
\end{titlepage}



\section{Introduction}

Nowadays the observation of the famous quantum Hall effect (QHE)~\cite{prange}
does not remain at the stage of the semiconductors, but it can be seen in
other materials. This is for example the case of
graphene, which is a projected
graphite of the group symmetry $C_3$ into two-dimensional spaces.
When such system is submitted to a perpendicular magnetic field,
it appears
an anomalous integer
Hall conductivity~\cite{novoselov,zhang} that is a manifestation of
the relativistic particle motions. It is originated
from different sources, which are the four-fold
spin/valley and the Berry phase due to the
pseudospin (or valley) precision when a massless
(chiral) Dirac particle exercises cyclotron motion.
This result was theoretically suggested by two groups~\cite{ando,sharapov}
independently.

The emergence of QHE in graphene 
during 2005 offered a laboratory for different investigations
and applications of some mathematical formalism. Many of them
brought from the QHE studies in semiconductors and used the Dirac operator
as the starting
point instead of the Landau Hamiltonian. Indeed, different questions
have been partially or completely solved but still some attentions to be paid
to others.
This concerns an eventual fractional quantum Hall effect (FQHE)~\cite{tsui82} in graphene
and related issues. 
Such interest can be linked to the fact that
FQHE is a fascinating subject because
starting from its appearance is still capturing
a great attention. This is due to 
its relations to different areas of
physics and mathematics.

There are many interesting theories have been appeared dealing with
some problems related to the anomalous FQHE, for instance
one may refer to the papers~\cite{yang2,goerbig,toke,khvesh}. In particular,
based on
different arguments
 some wavefunctions have been suggested as good candidates
to describe the subject~\cite{toke2,goerbig2,jellal}.
In fact, some of them used the composite fermion approach~\cite{cf}
and other made a straightforward generalization of the Halperin theory~\cite{halperin}
for non-polarized spin. However, some wavefunctions can be linked to each other
and recovered from a general proposal as has been seen in~\cite{jellal}.

Another theory was proposed by
the first author~\cite{jellal}
who studied the effect exhibited by
the relativistic particles
living on two-sphere $\mathbb{S}^2$ and submitted to
a magnetic monopole. In fact, he started by establishing a direct
connection between the Dirac and
Landau operators through the Pauli--Schr\"odinger
Hamiltonian $H_{\sf s}^{\sf SP}$.
This was helpful in the sense that
the Dirac eigenvalues and eigenfunctions were easily derived.
In analyzing $H_{\sf s}^{\sf SP}$ spectrum,
he showed that
there is a composite fermion nature supported by the presence of
two effective magnetic fields.
For the lowest Landau level (LLL), he argued that the basic physics of
graphene is similar to that of two-dimensional electron gas,
which is in agreement with the planar limit.
 For the higher Landau levels,
he proposed a $SU(N)$ wavefunction for different filling factors
that captures all symmetries. Focusing on the graphene case, i.e. $N=4$,
he gave different configurations
those allowed to recover some known results.

Motivated by different analysis concerning the anomalous FQHE
and in particular~\cite{jellal},
 we develop a real approach in order to describe
the subject. Taking advantage of our knowledge about the phenomena
in semiconductors, we introduce some effective interactions as relevant
ingredient in forming the Hamiltonian system. For this, we consider $M$-particles
in the presence of a perpendicular
magnetic field $B$ where the interaction is taken to be
two and three-body types. These have been successfully applied to
the conventional FQHE, for instance one may see~\cite{karabali,rao}.
Based on the connection between the Pauli--Schr\"odinger and Dirac Hamiltonian's,
we construct an appropriate Dirac operator that captures the basic feature
of the interacting system and where the Laughlin wavefunction~\cite{laughlin}
is its groundstate. 

Because of the basic physics is the same in both systems:  semiconductors and graphene
for the Laughlin series, we start by establishing the corresponding effective
potential. It will be generalized
to describe the composite fermions in graphene~\cite{castroneto} of filling factor
$\nu_{\sf cf}= 4\left(n+{1\over 2}\right)$, with $n=0, \pm 1, \pm 2, \cdots$.
This of course
captures the physics behind the fractional filling factors beyond Laughlin
series ${1\over 2l+1}$, with $l$ is integer, for FQHE in graphene.
Moreover, we return to the $SU(N)$ wavefunctions~\cite{jellal}
to firstly give a concrete example showing a particular type of potential.
Secondly, we show how the corresponding
filling factor can be linked to that generated from the composite fermions
pictures~\cite{castroneto}.
Different discusions will be reported elsewhere about the matter.

The present paper is organized as follows. In section $2$,
in order to show the difference between the interacting and free
particles cases, we start by
studying a system without interaction. We make use an approach
based on the
mapping between different Hamiltonian's to get
the corresponding spectrum and eigenfunctions. In section $3$, for the
later convenience, we give some discussions
about the Laughlin
potential that captures the interaction effect.
We introduce 
two and three-body interactions and analyze the
behaviour the new system in section $4$. In particular,
we show that the Laughlin wavefunction is a groundstate
of the present system and derive the spectrum's
for the excited states. We note that
the interactions
can be generated from a singular transformation
and allow us to establish a relation between
both Dirac systems:
free and interacting.
In section 5,
we consider the  composite Dirac fermions in terms of
an appropriate Jain potential, which can be obtained
by making use of straightforward generalization
of the Laughlin one.
In this framework, we analyze the $SU(N)$ wavefunctions and
show its basic features in section 6. Finally, we conclude and give
some perspectives of the present work.


\section{Dirac Hamiltonian}

We start
by developing our approach that will be subsequently generalized
to the case of the interacting system. This concerns
a mapping between different
spectrum's: Landau,  Pauli--Schr\"odinger and Dirac. Indeed,
let us consider one-relativistic particle living on the plane $(x,y)$
in presence of a perpendicular magnetic field $B$.
This can be described by
\begin{equation}\label{has}
H_{\sf PS}=\frac{1}{2m}\left[\vec{\sigma}\cdot\left(\vec{P}-\frac{e}
{c}\vec{A}\right)\right]^{2}
\end{equation}
where $\vec{\sigma}$ are the Pauli matrices and verify the usual
relations, namely
\begin{equation}\label{pm}
    \{\sigma_{i},\sigma_{j}\}=2\delta_{ij}, \qquad
    [\sigma_{i},\sigma_{i}]=2\epsilon_{ijk}\sigma_{k}.
\end{equation}
We will see how the Pauli--Schr\"odinger Hamiltonian $H_{\sf PS}$  can be used
to get the Dirac operator for the present system. Specifically,
it will be derived from the square root of $H_{\sf PS}$.
This mapping is useful in sense that the Dirac eigenvalues
and eigenfunctions will be easily obtained.

One way to establish the mentioned mapping is
to express $H_{\sf PS}$ in terms of the Landau
Hamiltonian, which describe free particle. This connection
has an important advantage because it leads to exactly derive
different quantities related to the subject.
Thus, by choosing the symmetric
gauge
\begin{equation}\label{gas}
    \vec{A}=\frac{B}{2}\left(-y,x\right)
\end{equation}
we show that $H_{\sf PS}$ can be mapped in the form
\begin{equation}\label{hasm}
H_{\sf PS}=\left(%
\begin{array}{cc}
  H_{\sf L} & 0 \\
  0 & H_{\sf L} \\
\end{array}%
\right)
-{B\over 2}\left(%
\begin{array}{cc}
  1 & 0 \\
  0 & -1 \\
\end{array}%
\right)
\end{equation}
where $H_{\sf L}$ is 
simply the Landau Hamiltonian and reads as
\begin{equation}\label{lhv}
H_{\sf L}=\frac{1}{2m}\left(\vec{P}-\frac{e}
{c}\vec{A}\right)^{2}
\end{equation}
which is extremely used in different contexts and in particular in
analyzing QHE in the semiconductor systems. To easiest way to resort
different spectrum is to write $H_{\sf L}$ in terms of the annihilation and creation
operators as
\begin{equation}
H_{\sf L} = {1\over 4}(a^\da a+ a a^\da )
\end{equation}
where, in the complex notation $z=x+iy$, $a$ and $a^\da$
are given by
\begin{equation}\label{ocra}
\hat{a}= 2{\pa\over \partial{z}}+\frac{B}{2}\bar{z}, \qquad
    \hat{a}^{\dag}=-2{\pa \over \partial{\bar{z}}}+\frac{B}{2}|z|^2.
\end{equation}
They verify the commutation relation
\begin{equation}\label{recom}
    [\hat{a}, \hat{a}^{\dag}]=2B.
\end{equation}
Since we are wondering to generalize different Hamiltonian's entering in
the game, let us convert $H_{\sf L}$ into its analytical form. 
It is not hard to obtain
\begin{equation}\label{hlc}
H_{\sf L}=-{1\over 2}\left\{ 4\frac{\partial^{2}} {\partial z_{i}\partial
\bar{z}_{i}}- B\left(z{\pa\over\partial{z}}-\bar{z}
{\pa\over\partial{\bar{z}}}\right)-{B^{2}\over4}\bar{z}\right\}.
\end{equation}
Hereafter, we set the fundamental constants
$\left(e, c, \hbar, m\right)$ to one. Clearly, the spectrum and
eigenfunctions of $H_{\sf PS}$ can be derived from that of
$H_{\sf L}$.


The above tools can be applied to analysis the anomalous QHE in
graphene. Indeed in such systems, the two-Fermi points, each with
a two-fold band degeneracy, can be described by a low-energy
continuum approximation with a four-component envelope
wavefunction whose components are labelled by a Fermi-point
pseudospin $=\pm1$ and a sublattice forming an honeycomb.
Specifically, the Hamiltonian for one-pseudospin component can be
obtained from (\ref{hasm}) under some consideration. This is
\begin{equation}\label{hadm}
H_{\sf D}={i\over \sqrt{2}} v_{\sf F}\left(%
\begin{array}{cc}
  0 & \hat{a}^{\dag} \\
  -\hat{a} & 0 \\
\end{array}%
\right).
\end{equation}
where $v_{\sf F}\thickapprox\frac{c}{100}$ is the Fermi velocity,
which will be set to one,  and
the many-body effects are neglected.
Its spectrum can be determined in a simple way if we introduce its
square. This is
\begin{equation}\label{hadsq}
    H_{\sf D}^{2}= {1\over 2}\left(%
\begin{array}{cc}
 \hat{a}^{\dag}  \hat{a}& 0 \\
  0 &  \hat{a} \hat{a}^{\dag}
\end{array}%
\right).
\end{equation}
It
is related to the Pauli-Schr\"{o}dinger
Hamiltonian (\ref{hasm}) up to some multiplicative constants. It
is clear that, $H_{\sf D}^{2}$ is written in terms of
the diagonal form of the Landau Hamiltonian (\ref{hlc}).
Therefore, it spectrum can easily be obtained.

We start by solving the eigenvalue equation
\begin{equation}\label{eigeqdr}
   H_{\sf D}^{2}\Psi=E_{\sf D}^{2}\Psi
\end{equation}
Since $H_{\sf D}^{2}$ has to do with the Landau
Hamiltonian (\ref{hlc}), the wavefunctions $\Psi$ should be
written in an appropriate form. This is
\begin{equation}\label{eigvec}
 \Psi_{m,n}=\left(%
\begin{array}{c}
 \psi_{m,n} \\
\psi_{m-1,n}
\end{array}%
\right)
\end{equation}
where the eigenfunctions $\psi_{m,n}$
are given by
\begin{equation}\label{eigfun}
    \psi_{m,n} (z,\bar z)=\frac{(-1)^{m}\sqrt{B^{m}m!}}{\sqrt{2^{n+1}
\pi(m+n)!}}z^{n}L^{n}_{m}(\frac{z\bar{z}}{2})e^{-\frac{B}{4}z\bar{z}},\qquad
    m,n=0,1,2\cdots.
\end{equation}
The corresponding Landau
levels are given by
\begin{equation}\label{landlev}
    E_{{\sf D}}^{2}(m)=Bm.
\end{equation}
We can show that the normalized eigenfunctions
of $H_{\sf D}$ take the form
\begin{equation}\label{eigfuno}
    \Psi_{m\neq0,n}=\frac{1}{\sqrt{2}}\left(%
\begin{array}{c}
  \psi_{|m|,n} \\
-\,i\,\psi_{|m|-1,n} \\
\end{array}%
\right)
\end{equation}
with the convention $\psi_{-1,n}=0$. Note that, the zero mode
wavefunction is
\begin{equation}\label{eigfunf}
    \Psi^{(0,n)}=\left(%
\begin{array}{c}
  \psi_{0,n}\\
0
\end{array}%
\right)\cdot
\end{equation}
Their energy levels read as
\begin{equation}\label{enelev}
    E_{{\sf D}} (m)=\pm \sqrt{B|m|}.
\end{equation}
This makes a difference with respect to the Landau spectrum for the
present system. In fact, it
has no zero energy and is discrete as well, unlike~(\ref{enelev}).
These what make the integer QHE is an unconventional effect
in graphene.

The above analysis can easily be generalized to
many-body system without interaction.
This leads to a spectrum as sum over all
single particle ones and eigenfunctions
as tensor products.
However, we will give a generalization of the system by
introducing a kind of interaction that is behind
FQHE in semiconductors and see what is its
influence on graphene.


\section{Laughlin wavefunction}

To develop our main idea we need first to start
 from what we know so far about the Laughlin wavefunction.
This latter is involving a Jastrow factor that is
originated from a specific interaction  between particles.
In fact, it corresponds to an
artificial model that captures the basic physics of the filling factor
$\nu={1\over 2l+1}$. This is an typical example
to get more general results, which concern the composite
fermions as well as the $SU(4)$ wavefunctions for graphene. For our task it
is necessary to start from the Laughlin theory for FQHE
in graphene.

To talk about the Laughlin series for FQHE, it is convenient
to introduce some discussions about LLL because it is a rich
level and has many interesting properties. Indeed, in such level the Landau system
behaves like a non-commutative one, which is governed
by the commutation relation
\begin{equation}\label{nonr}
[z, \bar z]_{\sf LLL}= {2\over B}.
\end{equation}
This effect can be interpreted as follows. In the present level,
the potential energy is strong enough than kinetic energy and
therefore the particles are glue in the fundamental level.
Using the same analysis as for the case of $H_{\sf L}$,
is not hard to get a basis as set of the eigenstates. They are~\cite{jellal2}
\begin{equation}\label{llleg}
|m\rangle_{\sf LLL} = 
{1\over \sqrt{m!}} (\hat a^\da|_{\sf LLL})^m|0\rangle_{\sf LLL}.
\end{equation}
where the annihilation and creation operators reduce now to
\begin{equation}\label{reaap}
\hat a|_{\sf LLL}= {B\over 2} z, \qquad \hat a^\da|_{\sf LLL}= {B\over 2} \bar z.
\end{equation}
The projection into the complex plane leads to the
eigenfunctions
\begin{equation}\label{lllb}
\Psi_m{(z,\bar z)} =  \sqrt{B^{m+1}\over 2^{m+1}\pi m!}z^m
e^{-{B\over 4}|z|^2}.
\end{equation}
As we will see very soon, this can be generalized to get
more interesting results. In particular, these will offer a way
to get in contact with the Laughlin theory~\cite{laughlin} for FQHE.

Returning now to discuss FQHE in graphene. In doing so,
Let us consider $M$-particles in LLL, which of course means that all $m_i=0$
with $i=1,\cdots,M$ and each $m_i$ corresponds to
the spectrum~(\ref{eigfuno}--\ref{enelev}).  The total wavefunction
of zero-energy Landau level~(\ref{eigfunf})
can be written in terms of the Slater determinant. This is
\begin{equation}\lb{nwps}
\psi(z,\bar{z})= \epsilon^{i_1 \cdots i_M} z_{i_1}^{m_1} \cdots z_{i_M}^{m_M}
\exp\left(-{B\over 4}\sum_i^M|z_i|^2\right)
\end{equation}
where $\epsilon^{i_1 \cdots i_M}$ is the fully
antisymmetric tensor and ${m_i}$ are integers. It is relevant to write this
wavefunction as Vandermonde determinant. We have
\begin{equation}\lb{nwp}
\psi(z,\bar{z})= {\mbox{const}}\,  \prod_{i<j}^M\left(z_i-z_j\right)
\exp\left(-{B\over 4}\sum_i^M|z_i|^2\right).
\end{equation}
The corresponding energy is \beq E(m_i)=B\left(\sum_i m_i
+{M}\right). \eeq This can be interpreted by remembering the
Laughlin wavefunction
\begin{equation}\lb{lw}
\psi_{\sf Laugh}^l(z,\bar{z})=  \prod_{i<j}^M
\left(z_i-z_j\right)^{2l+1}\exp\left(-{B\over 4}\sum_i^M|z_i|^2\right).
\end{equation}
It is well-known that it has many interesting features and good ansatz to describe
the fractional QHE at   the filling factor $\nu ={1\over 2l+1}$,
with $l$ is integer value. It is clear that~(\ref{nwp})
is nothing but the first Laughlin state that corresponds to $\nu=1$. Actually,~(\ref{nwp})
is describing the first quantized Hall plateau of the integer QHE.
It is also convenient for us to write~(\ref{lw})
under another form. This is
Note that~(\ref{lw}) can also be written as
\begin{equation}\lb{lw2}
|l\rangle_{\sf Laugh}= \left\{\epsilon^{i_1 \cdots i_M} z_{i_1}^{m_1}
\cdots z_{i_M}^{m_M}\right\}^{2l+1}
|0\rangle.
\end{equation}
Consequently, the wavefunction for the particles in the graphene LLL
are identical to those of the 2DEG LLL. We conclude that
the basic physics in both systems is the same. More
discussion about this issue can be found in~\cite{toke}.

As far as we know FQHE is a consequence of the interacting particles
in the presence of an external magnetic field. Then, obviously
the Laughlin wavefunction is a manifestation of a specific
interaction type, which coded in the Jastrow factor.
At this point, one may ask how does look like such
interaction and to reply this question, let us write~(\ref{lw})
as
\begin{equation}\lb{lw3}
\psi_{\sf Laugh}^l(z,\bar{z})=  F^{\sf Laugh}(z,\bar{z})\exp\left(-{B\over 4}\sum_i^M|z_i|^2\right)
\end{equation}
where the function $F^{\sf Laugh}(z,\bar{z})$ ant its anti-holomorphic partner are given by
\begin{equation}\lb{jla}
 F^{\sf Laugh}(z,\bar{z})= \prod_{i<j}^M
\left(z_i-z_j\right)^{2l+1}, \qquad
\bar F^{\sf Laugh}(z,\bar{z})=
\prod_{i<j}^M
\left(\bar z_i-\bar z_j\right)^{2l+1}.
\end{equation}
These are nothing but the results of a specific interaction, which
generated from particles. To clarify this point, let us
define two functions as the derivatives with respect to the variables $z$ and
$\bar z$ of $F^{\sf Laugh}$ and $\bar F^{\sf Laugh}$, respectively. They are
\begin{equation}\lb{pcoin}
(V_z^{\sf Laugh})_i= -2i{\pa\over \pa z_i} (\ln  F^{\sf Laugh} -\ln\bar F^{\sf Laugh}), \qquad
(V_{\bar z}^{\sf Laugh})_i = 2i{\pa\over \pa \bar z_i} (\ln  F^{\sf Laugh} -\ln\bar F^{\sf Laugh}).
\end{equation}
Explicitely, the corresponding potential can be written as
 \begin{equation}\lb{pcoinz}
(V_z^{\sf Laugh})_i= -2i(2l+1) \sum_{j\neq i}{1\over z_i -z_j}, \qquad
(V_{\bar z}^{\sf Laugh})_i = -2i (2l+1) \sum_{j\neq i}{1\over \bar z_i -\bar z_j}
\end{equation}
which is resulting from of interaction of the $i^{th}$ particle
with the remaining $j$ ones.
These can also be identified to the components of the Chern--Simon potential,
which has been employed at many occasions to study the conventional QHE. For
instance, more discussion about such identification
can be found in~\cite{rao}. This of course open
another way to describe the anomalous QHE by making use the field theory
approach.
Strictly speaking the present potential is, in reality,
a part of an Hamiltonian whose groundstate
is exactly the Laughlin wavefunction. Moreover, we show that
how (\ref{pcoin}) can be generalized and used to
discuss other filling factors beyond Laughlin states for FQHE in graphene.
This mainly we will be done by adopting the composite fermion picture~\cite{cf}
and using the $SU(4)$ wavefunctions~\cite{jellal} as an illustration.

\section{Many-body interactions}

As claimed before we are wondering to investigate
 the basic feature of FQHE in graphene.
In doing so, we have to take account of a kind of interaction. For
this, we start with a Hamiltonian that has been considered at
different occasions and is relevant to describe FQHE
at the filling factor $\nu={1\over 2l+1}$.
Further, we show that this will be applied to graphene and
allow us to derive different results.
Mainly, we apply the approach seen before
for free relativistic particles to achieve
our goal. In fact, we start from a Hamiltonian
generalizing the Landau one to map a bridge between
the Pauli--Schr\"odinger and Dirac Hamiltonian's
for interacting particles. Of course,
this will be done by constructing the appropriate
annihilation and creation operators those will be used
to build the excited states and derive the corresponding
energies.

\subsection{Excited states}

In the beginning, we start by resorting the spectrum
for a system that include a specific interaction.
This can be done by constructing
an explicit model that has the Laughlin wavefunction as an
exact groundstate.
Let us particles living on the plane in
the presence of an external magnetic field and
involve some kind of interactions.
 When these are of two-body and three-body
types, the present system can be described
by the Hamiltonian
\begin{eqnarray}\label{Hamro}
 H_{\sf L}^{\sf int} &=& \frac{1}{2}\sum^{M}_{i}\left(p_{i}-A_{i}\right)^{2}+{\eta}
\sum^{M}_{i,j\neq i}\frac{1}{(z_{i}-z_{j})}
\left(2\frac{\partial}{\partial\bar{z}_{i}}- {i} A_{z}\right)-
{\eta}\sum^{M}_{i,j\neq i}\frac{1} {(\bar{z}_{i}-\bar{z}_{j})}
\left(2\frac{\partial}{\partial z_{i}}-{i} {A}_{\bar z}\right)
\nonumber\\
 &&
+{2\eta^{2}}\sum^{M}_{i,j\neq i,k\neq
i}\frac{1}{(z_{i}-z_{j})(\bar{z}_{i}-\bar{z}_{k})}
\end{eqnarray}
where the coordinate $z_{i}=x_{i}+iy_{i}$ denote the position of the
$i^{th}$ particle and $\eta$ is the inverse of the Laughlin series,
i.e. $\eta=2l+1$ is an odd integer. The second and
third terms in (\ref{Hamro}) are two-body interactions whereas the
fourth one is a three-body type. In the symmetric gauge
where the
complex components of the vector potential $\vec{A}$ are
\begin{equation}\label{jaujs}
A_{z}= - {i\over 2} Bz, \qquad {A}_{\bar z}={i\over 2} B \bar z
\end{equation}
the Hamiltonian (\ref{Hamro}) takes the form
\begin{eqnarray}\label{hamr}
  H_{\sf L}^{\sf int} &=& -{1\over 2}\sum^{M}_{i}\left\{4\frac{\partial^{2}}
{\partial z_{i}\partial \bar{z}_{i}}-
B\left(z_{i}\frac{\partial}{\partial
z_{i}}-\bar{z}_{i}\frac{\partial}{\partial \bar{z}_{i}}\right) -
{B^2\over4} z_{i}\bar{z}_{i}\right\} + \eta  \sum^{M}_{i,j\neq i}
\frac{1}{(z_{i}-z_{j})}\left(2\frac{\partial}{\partial\bar{z}_{i}}-\frac{B}{2}
z_i\right)
\nonumber\\
 &&
  - \eta  \sum^{M}_{i,j\neq i}\frac{1}{(\bar{z}_{i}-\bar{z}_{j})}
\left(2\frac{\partial}{\partial
    z_{i}}+\frac{B}{2} \bar z_i\right)+ 2\eta^{2}  \sum^{M}_{i,j\neq i,k\neq
i}\frac{1}{(z_{i}-z_{j})(\bar{z}_{i}-\bar{z}_{k})}.
\end{eqnarray}
It is obvious that
when $\eta$ is switched off we get the Landau Hamiltonian for
a free system of $M$-particles. Therefore, the presence of this parameter will
play an important role in discussing different issues related
to the anomalous FQHE.

To analyze the behaviour of interacting relativistic particles,
it is relevant
to derive the corresponding Dirac Hamiltonian. As far as we know,
the easiest way to do is to adopt the technical used in section 2,
which based on the mappings between three Hamiltonian's.
More specifically, this can be done by making use of an algebraic
method that requires to introduce the appropriate annihilation and
creation operators. Indeed, in terms of the interaction
they can be realized as
\begin{equation}\label{opcr}
\hat{A}_{i}=2{\pa\over\pa{z_{i}}}+\frac{B}{2}\bar{z}_{i}-2\eta \sum^{M}_{j\neq
i}\frac{1}{z_{i}-z_{j}},\qquad
\hat{A}^{\dag}_{i}=-2{\pa\over\pa_{\bar{z}_{i}}}+\frac{B}{2}z_{i}-2\eta\sum^{M}_{j\neq
i}\frac{1}{\bar{z}_{i}-\bar{z}_{j}}.
\end{equation}
It is not hard to show the commutation relation
\begin{equation}\label{comre}
    \left[\hat{A_{i}},\hat{A}^{\dag}_{j}\right]=2B\delta_{ij}
    -4\eta \sum_{i\neq j}^{M} \delta(z_{i}-z_{j}) - 4\eta \sum_{i\neq j}^{M}
\delta(\bar{z}_{i}-\bar{z}_{j})
\end{equation}
where for instance the second term is resulted by acting
$\pa_{z_{i}}$ on the third term of $\hat{A}^{\dag}_{j}$ and so on.
These allow us to get a diagonal form of $H_{\sf L}^{\sf int}$, such as
\begin{equation}\label{hamcr}
    H_{\sf L}^{\sf int}={1\over 2}
\left(\sum^{M}_{i}\hat{A}^{\dag}_{i}\hat{A}_{i}+ MB\right)-2\eta
\sum_{i\neq j}^{M} \delta(z_{i}-z_{j}).
\end{equation}
Since $\delta( z_{i}- z_{j})$ is irrelevant when acting on an
antisymmetric wavefunction, it can be ignored. Moreover,
$\hat{A}_{i}$ and $\hat{A}^{\dag}_{i}$ satisfy the canonical
harmonic oscillator commutation rules even for non-zero $\eta$.
Hence, the Hamiltonian is exactly soluble for all the excited
states. A basis of wavefunctions describing excitations of the
$\nu=\frac{1}{\eta}$ states in the Laughlin theory is therefore
given by
\begin{equation}\label{wvelaug}
    |\{m_{i}\},\eta \rangle_{\sf L}=\epsilon^{i_{1}\cdots i_{M}}\hat{A}^{\dag^{m_{1}}}_{i_{1}}\cdots
    \hat{A}^{\dag^{m_{M}}}_{i_{M}}\left(\epsilon^{i_{1}\cdots i_{M}}\hat{A}^{\dag^{0}}_{i_{1}}.
    \hat{A}^{\dag^{M-1}}_{i_{M}}\right)^{\eta-1}|0\rangle
\end{equation}
where $m_{i}$ are integers such that $0\leq m_{1}<\cdots<m_{M}$.
The corresponding eigenvalues are
\begin{equation}\label{entot}
 E_{\sf L}^{\sf int}(\left\{m_{i}\},\eta\right) = B\left(
\sum_{i}^{M}m_{i}+ \frac{1}{2}M\left[ (\eta-1)(M-1)+1\right]\right).
\end{equation}
For later convenience, let us project
these states on the complex plane. Indeed, by requiring
that $m=i-1$, we end up with the wavefunction
\begin{equation}\lb{exlw}
\psi_{\sf L}^l(z,\bar{z})=  \prod_{k<l}^M \left(z_k-z_l\right)
\prod_{i<j}^M
\left(z_i-z_j\right)^{2l+1}\exp\left(-{B\over 4}\sum_i^M|z_i|^2\right).
\end{equation}
An interesting result can immediately be noticed. Indeed, now
it is obvious to see
that the Laughlin wavefunction is an exact groundstate of
(\ref{hamr}) and corresponds to the energy ${MB\over 2}$.
More precisely, it is easy to check
that $\hat{A}_{i}$ annihilates
$|\eta\rangle_{\sf Laugh}\equiv |l\rangle_{\sf Laugh}$,
namely
\begin{equation}\label{LLL}
    {\hat{A}}_{i} |\eta\rangle_{\sf Laugh}=0, \qquad \forall i.
\end{equation}
This is nothing but the lowest Landau level condition.
It is clear that for $\eta=1$ we recover the fully occupied state
seen before. This show that the present model captures some basic features
of the phenomenon.

In the next, we will show how the above tools can be used to describe
FQHE in graphene. In the beginning we start by establishing the appropriate
background to do our task. In particular, we determine the Dirac
Hamiltonian and related matters for the present system.

\subsection{Spinors}

To apply the technical used before, it is convenient to
write down the Pauli--Schr\"odinger Hamiltonian
$ H_{\sf PS}^{\sf int}$ for interacting
particles in the presence of $B$. The easiest way to do
is to give its expression  in terms of the Hamiltonian $H_{\sf L}^{\sf int}$.
The advantage of this mapping is make use
an algebraic method to diagonalize and therefore derive
the associate spectrum. This will be helpful in establishing
the corresponding Dirac formalisms.

One way to write down $ H_{\sf PS}^{\sf int}$
is to generalize the Hamiltonian~(\ref{hasm}) for free system
to the interaction case. In doing so, we
get the form
 \begin{equation}\label{pshami}
   H_{\sf PS}^{\sf int}=\left(%
\begin{array}{cc}
  H_{\sf L}^{\sf int} & 0 \\
  0 & H_{\sf L}^{\sf int}\\
\end{array}%
\right)+ \left(%
\begin{array}{cc}
  2\eta\sum_{i\neq j}^{M}
\delta(z_{i}-z_{j}) & 0 \\
  0 &
-2\eta\sum_{i\neq j}^{M}
\delta(\bar{z}_{i}-\bar{z}_{j})\\
\end{array}%
\right) -{1\over 2} MB\left(%
 \begin{array}{cc}
  1  & 0 \\
  0 & -1 \\
\end{array}%
\right).
\end{equation}
This is actually involving three matrices and the second
one is particular.  On the other hand, it can
be identified to that used by Koshino and Ando~\cite{ando2}
in discussing the diamagnetism in disordered graphene
in arbitrary magnetic fields.
However, in our case it be ignored since does not affect the
antisymmetric wavefunction.


To start talking about different issues, we need first
to resort the spectrum as well as the eigenstates
of $ H_{\sf PS}^{\sf int}$. Clearly, these matters can be
easily derived to get
the excited states as spinors
\beq
|\{m\},\eta \rangle_{\sf PS}=
\left(\begin{array}{c}
  |\{m_{i}\},\eta \rangle_{\sf L}\\
|\{m_{i}-1\},\eta \rangle_{\sf L}
\end{array}\right)
\eeq
where the eigenstates $|\{m_{i}\},\eta \rangle_{\sf L}$
are those for the Hamiltonian (\ref{hamcr}), with
the convention $|\{-1\},\eta \rangle$ is a null state.
The corresponding eigenvalues are
\begin{equation}\label{valpd}
    E_{\sf PS}^{\sf int}(m,\eta)=B\left(\sum_{i=1}^{M}m_{i}+\frac{1}{2}(\eta-1)M(M-1)\right).
\end{equation}
We have some remarks in order. For $m_i=0$ and $\eta =1$, there is a zero
mode energy that corresponds to a fundamental state. However, for the case
where  $m_i=0$ and $\eta \neq 1$, we end up with
\begin{equation}\label{valpd0}
    E_{\sf PS}^{\sf int}(m=0,\eta)=\frac{1}{2}(\eta-1)BM(M-1).
\end{equation}
This captures the basic features of the excited states for the
present system, which is exactly the LLL condition. It offer another discussion about
spin as a degree of freedom and its influence on the discussion
reported before. Specifically, its relation with the conventional QHE,
but this is out of scope of the present work.

At this level
more discussion can be reported on the spectrum and in particular
the corresponding Hilbert space, which is a set of all states
$|\{m_{i}\},\eta \rangle_{\sf PS}$. It is invariant under
the semi-direct group $G=U(1)\times \mathbb{C}$ and therefore
for an unitary projective representation $T(g)$ of $G $ we have the
invariance property
\begin{equation}\label{inva}
T(g)H_{\sf PS}^{\sf int} = H_{\sf PS}^{\sf int} T(g), \qquad \forall g\in G
\end{equation}
where an element of $G$ is a $2\times 2$ matrix of the form
\begin{equation}\label{gelm}
g=\left(
\begin{array}{cc}
  a  & b \\
  0 & 1 \\
\end{array}%
\right).
\end{equation}
This representation can be further explored and then
make a group theory approach to
investigate the basic features of the present system.
It
is acting on the spinors trough the relation
\begin{equation}\label{rasp}
T(g)|\{m\},\eta \rangle_{\sf PS}= j(g,z) g^{-1}|\{m\},\eta \rangle_{\sf PS}
\end{equation}
 where the quantity $g^{-1}|\{m\},\eta \rangle_{\sf PS}$
is the pull-back of $|\{m\},\eta \rangle_{\sf PS}$ through the mapping
$z \mapsto g\cdot z$ such as
\beq
g\cdot z= az+b.
\eeq
The function $j(g,z)$ is and automorphic factor. One also can add more analysis
about the incompressibility 
in terms of the area preserving diffeomorphism algebra that
can be realized in terms of the ingredients of the
present system.

\subsection{Interacting Dirac particles}

The above mathematical background can be applied to analyze FQHE in graphene. For this,
it is word while to establish a Dirac formalism for the interacting particles. One way to do this
is to make an analysis based on the former approach for non-interacting particles. This requires
an analogue Dirac operator in terms of the annihilation and creation operators.

Right now we have all materials needed to do our job. Indeed, the appropriate
interacting operator can be written in similar way
to (\ref{hamd}). Thus for the $i^{th}$ particle, we have
\begin{equation}\label{hamd}
    (H_{\sf D}^{\sf int})_i={i\over \sqrt{2}}\left(%
\begin{array}{cc}
  0 & \hat{A}_{i}^{\dag} \\
  -\hat{A}_{i} & 0 \\
\end{array}%
\right)
\end{equation}
where $\hat{A}_{i}$ and $\hat{A}_{i}^{\dag}$ are given in (\ref{opcr}).
To make a comparison with the Pauli--Schr\"odinger Hamiltonian, let us
define the square of $H_{\sf D}^{\sf int}$. This is given by
\begin{equation}\label{hamds}
    \left[\left(H_{D}^{\sf int}\right)_i\right]^{2}={1\over 2}\left(%
\begin{array}{cc}
 \hat{A}_{i}^{\dag}  \hat{A}_{i}& 0 \\
  0 &  \hat{A}_{i}\hat{A}_{i}^{\dag} \\
\end{array}%
\right).
\end{equation}
Summing over all $M$-particles
and if we forget about
all constants entering in different Hamiltonian's,
it is not hard to end up with the relation
 \begin{equation}\label{hamild2}
  \left(H_{\sf D}^{\sf int}\right)^{2} =H_{\sf PS}^{\sf int}.
\end{equation}
This equivalence shows that how the applied
approach is relevant as well to analyze the
interacting particles. Remember that $\del(z_i-z_j)$ and $\del(\bar z_i-\bar z_j)$
can be dropped because they do not affect the antisymmetric wavefunction.
Therefore, we can carry out our previous
study concerning free Dirac particles and its connection
to FQHE on graphene to the present case and underline the difference.

It is clear that, $\left(H_{\sf D}^{\sf int}\right)^{2}$
is written in terms of the
diagonal form of the Hamiltonian (\ref{Hamro}) and therefore its
spectrum can easily be obtained.
Then, the eigenstates of $H_{\sf D}^{\sf int}$ take the
form
\begin{equation}\label{funddin}
|\{m\}\neq 0,\eta \rangle_{\sf D}=
\left(\begin{array}{c}
 |\{|m_{i}|\},\eta \rangle_{\sf L}\\
- i |\{|m_{i}|-1\},\eta \rangle_{\sf L}
\end{array}\right)
\end{equation}
and the zero mode is given by
\begin{equation}\label{funddin0}
|\{0\},\eta \rangle_{\sf D}=
\left(\begin{array}{c}
|\{0\},\eta \rangle_{\sf L}\\
0
\end{array}\right).
\end{equation}
The corresponding eigenvalues are
\begin{equation}\label{valpdi}
    E_{\sf D}^{\sf int}(m_i,\eta)=\pm
\sqrt{B\left|\sum_{i=1}^{M}m_{i}+\frac{1}{2}(\eta-1)M(M-1)\right|}.
\end{equation}
It can be projected on LLL that is realized by requiring that $m_i=0$.
This operation
leads to the lowest energy for the excited Dirac states, such as
\begin{equation}\label{valpdi0}
    E_{\sf D}^{\sf int}(m_i=0,\eta)=\pm
\sqrt{\frac{1}{2}(\eta-1)M(M-1)}.
\end{equation}
Clearly, for $\eta =1$ we get a zero energy mode. This is
an agreement with the fully occupied state $\nu=1$ see before
fore the relativistic particles.

To close this part, one can also report some discussions about the symmetry
group that leaves invariance the Dirac Hamiltonian for
the present case. These of course will offer a theory
group approach to analyze the interacting relativistic
particles.

\subsection{Gauge transformation}

Sometimes it is relevant to make the appropriate transformations
in order to get some information about a system
under consideration. This has an interst in
sense that we can derive the integrability of the
system from its old partner. For this,
we are going to establish a concrete relation between the free
and interacting
Dirac operators. This can be done by
considering an appropriate singular gauge transformation
that captures the basic features the interacting term involved in
different Hamiltonian's see before.
It can also be seen as a kind of bosonization
of the present system, for this we may refer to the
reference~\cite{karabali}.
The relation can be served as
a good candidate to test the previous analysis concerning
the interacting particle case.

We start with a system of $M$-relativistic fermions of mass
$m$ in the presence of a external magnetic field. Without
interaction, this system is described by the total Dirac Hamiltonian
in the complex coordinates, such as
\begin{equation}\label{hd}
    H_{\sf D}={i\over \sqrt{2}}\left(%
\begin{array}{cc}
  0 &
 \sum_i^M \left(-2\frac{\partial}{\partial \bar z_{i}}+\frac{B}{2} {z}_{i}\right) \\
-\sum_i^M \left(2\frac{\partial}{\partial z_{i}}+\frac{B}{2} \bar{z}_{i} \right)& 0 \\
\end{array}%
\right).
\end{equation}
This of course an immediate generalization of the Hamiltonian for
one free particle. Therefore, its spectrum is nothing but
a summation over that of one particle and the wavefunctions
are tensor product of one single eigenfunctions.

To explicitly determine the element that governs such transformation,
we adopt the method used by Karabali and Sakita~\cite{karabali}. In doing so,
let us consider two wavefunctions 
and establish a link between them. This can be done
by considering a singular gauge transformation, such as
\begin{equation}\label{trajau}
    \Psi(z_{1},\cdots,z_{N})=U
\Phi(z_{1},\cdots,z_{N})
\end{equation}
where $\Psi$ and $\Phi$ are, respectively, totally antisymmetric
and symmetric wavefunctions. This transformation can be identified to
an element of the unitarty group $U(N)$ and should be written
in terms of a suitable function $\Theta$. This is
\begin{equation}\label{opuni}
    U=e^{-i\Theta(z_{1},\cdots,z_{N})}.
\end{equation}
 Clearly from (\ref{trajau}) we have some information about $\Theta$. Indeed,
it should be defined such that
 an interchange of a
pair of variables gives the constraint
\begin{equation}\label{inter}
\exp\left[{-i\Theta(z_{1}, \cdots, z_i, \cdots, z_j, \cdots,z_{N})}\right]
=-\exp\left[{-i\Theta(z_{1}, \cdots, z_j, \cdots, z_i, \cdots,z_{N})}\right].
\end{equation}
Consequentely,  $\Theta$ should be realized in such way that
(\ref{inter}) must be fullfiled. Thus, we map $\Theta$
in terms of our language as
\begin{equation}\label{realis}
\Theta(z_{1},\cdots,z_{N})=\eta \sum_{i<j}{\rm Im}
\ln({z}_{i}-{z}_{j}).
\end{equation}
This function  is not only interesting in the present case but
also in the field theory. In fact, it can be linked to
the relative angle between the position vectors the
$i^{th}$ and $j^{th}$ particles. This is
\begin{equation}\label{relan}
\Theta(z_{1},\cdots,z_{N})= (\eta -1)\sum_{j\neq i}^{M}\te_{ij}
\end{equation}
where $\te_{ij}$ is related to the Chern-Simons
potential of $i^{th}$ particle through the relation
\begin{equation}\label{apcs}
A_i^{\sf CS} = (\eta -1)\vec{\nabla}_i \sum_{j\neq i}^{M}\te_{ij}=
(\eta -1) \sum_{j\neq i}^{M}{\vec e_z \times (\vec r_i -\vec r_j)
\over |\vec r_i -\vec r_j|^2}
\end{equation}
which was used to study the anyon systems.
This mapping can offer another way to talk about the
interacting particles in terms of the Dirac formalism. Also
one could think to a make matrix model analysis
of the present system by adopting the Susskind--Polychronakos
approach~\cite{susskind,poly}.

Now we have all ingredients to do our task. Indeed,
since (\ref{trajau}) is a singular gauge
transformation, the Hamiltonian for $\Phi$ can be obtained from
(\ref{hd}) for $\Psi$. Indeed, we show that
\begin{equation}\label{hdint}
    H_{\sf D}^{\sf int}=U^{\dag}H_{\sf D}U.
\end{equation}
With this, we are now able to generate a specific
potential that governs the interaction between particles.
More precisely,
is not hard to end up with the Dirac Hamiltonian
for interacting particles in the presence of a magnetic
field, which is
\begin{equation}\label{hamd1}
    H_{\sf D}^{\sf int}={i\over \sqrt{2}}\left(%
\begin{array}{cc}
  0 &
  \sum_i^M \left(-2{\pa\over \partial_{\bar{z}_{i}}} +\frac{B}{2}z_{i}\right)
-2\eta\sum^{M}_{i,j\neq i}\frac{1}{\bar{z}_{i}-\bar{z}_{j}}  \\
-\sum_i^M \left(2{\pa\over \partial_{z_{i}}} + \frac{B}{2}\bar{z}_{i}\right)
+2\eta \sum^{M}_{i,j\neq i}\frac{1}{z_{i}-z_{j}} & 0 \\
\end{array}%
\right).
\end{equation}
This form shows how the gauge transformation can be employed
to produce some kind of interaction from the free case.
Moreover,
the above mapping tells us that the spectrum
and corresponding
eigenfunctions of $H_{\sf D}^{\sf int}$ can be deduced from those
of $H_{\sf D}$. For instance, we have a similar relation
to~(\ref{trajau}) between the states $|{m_i},{n_i}\rangle$ for $M$-particles
those are tensor products of single states~(\ref{eigfun}-\ref{eigfunf})
and $|\{m_i\},\eta\rangle_{\sf D}$.

We conclude this part by noting that the involved gauge
transformation can be generalized to higher order in terms
of some interactions. This can be done by considering more that
three-body correlations and therefore offers different
integrability of some models. This will be interesting in the
FQHE world because recall that this is based on such interactions
to be understood.

\section{Composite fermion wavefunction}

To give an explanation of the filling factors beyond the
Laughlin states, Jain has introduced the composite fermion
formalism~\cite{cf}. In fact, they
 are new kind of particles appeared in
condensed
matter physics to provide an explanation of the behavior of particles
moving in the plane when a strong magnetic field $B$
 is present. Particles
possessing $2p$, with $p\in N^*$, flux quanta (vortices) can be
thought of being composite fermions. One of the most important features of
them is they feel effectively a magnetic field of the form
\beq
\lb{cfm}
B^*=B \pm 2p\Phi_0\rho
\eeq
where in our convention the unit flux is $\Phi_0= 2\pi$.
Recalling the relation between the filling factor $\nu$ and $B$,
one can define a similar quantity for the field $B^*$. This can be written as
\beq\lb{defi}
\nu^* = 2\pi {\rho\over B^*}.
\eeq
It is clear that from (\ref{cfm}), the factors $\nu$ and
$\nu^*$ can be linked to each other through
\beq\lb{rnns}
\nu ={\nu^*\over 2p\nu^* \pm 1}.
\eeq
This relation has been used to deal with different issues
related to QHE in 2DEG. More discussions about the mapping
 (\ref{rnns}) and its applications can be found
in~\cite{cf}.

As an immediate consequence of (\ref{rnns}), we can map
the anomalous FQHE in terms of IQHE in graphene. Indeed,
it is easy to see
\beq\lb{nunus}
\nu^{\sf G}= 2{2n+1\over 4p(2n+1) \pm 1}.
\eeq
This result has been obtained in different contexts,
for instance see~\cite{castroneto}. It is obvious that for
$n=0$, we obtain Laughlin states and the anomalous IQHE
can be recovered by fixing $p=0$. Moreover, (\ref{nunus})
tells us that the Jain's series is quite different from the
2DEG. Therefore, it is interesting to focus on FQHE in graphene.

Remembering the Laughlin potential established before,
it is natural to ask about that behind  $\nu^{\sf G}$.
To reply this question, we build
a wavefunction corresponding to $\nu^{\sf G}$ in terms of
a generalized potential.
 In the beginning, let us
see how CF picture does look like for the Laughlin states,
which are the same for both systems: semiconductors and graphene. Indeed,
following Jain idea~\cite{cf}, one can write the Laughlin wavefunction
as
\begin{equation}\lb{lwcf}
\psi_{\sf Laugh}^p(z,\bar{z})=  \prod_{i<j}^M
\left(z_i-z_j\right)^{2p}\chi_1\exp\left(-{B\over 4}\sum_i^M|z_i|^2\right)
\end{equation}
where $\chi_1$ is the wavefunction of the fully occupied state $\nu=1$.
Now it is clear that for higher Landau level, one may generalize~(\ref{lwcf})
to that of the form
\begin{equation}\lb{lwcfn}
\psi_{\sf Jain}^{p,n}(z,\bar{z})=  \prod_{i<j}^M
\left(z_i-z_j\right)^{2p}\chi_{2(2n+1)}\exp\left(-{B\over 4}\sum_i^M|z_i|^2\right)
\end{equation}
and $\chi_{2(2n+1)}$ is describing the composite fermions
at the anomalous
integer filling factor $\nu_{\sf cf}$. This is a obvious mapping between the particles
generating FQHE in graphene and composite fermions behind unconventional
integer QHE.

At this level, one may ask about the potential of interaction
that is responsible for the emergence of  $\nu^{\sf G}$
and therefore the model that describes such case.
One way to answer
this inquiry is to make a straightforward generalization of the Laughlin
potential~(\ref{jla}-\ref{pcoin}). Then by analogy, we can define
the Jain potential as
\begin{equation}\lb{pcoinj}
(V_z^{\sf Jain})_i= -2i{\pa\over \pa z_i} (\ln  F^{\sf Jain} -\ln\bar F^{\sf Jain}), \qquad
(V_{\bar z}^{\sf Jain})_i = 2i{\pa\over \pa \bar z_i} (\ln  F^{\sf Jain} -\ln\bar F^{\sf Jain}).
\end{equation}
Clearly, for $n=0$ we recover~(\ref{jla}-\ref{pcoin}). Thus, we can write
the Jain wavefunction in terms of the function $F^{\sf Jain}(z,\bar{z})$ as
\begin{equation}\lb{jwfg}
\psi_{\sf Jain}^{p,n}(z,\bar{z})=  F^{\sf Jain}(z,\bar{z})\exp\left(-{B\over 4}\sum_i^M|z_i|^2\right).
\end{equation}
Of course it should be described by some Hamiltonian that
involves $(V_z^{\sf Jain})_i$ and $(V_{\bar z}^{\sf Jain})_i$
as important ingredients. This requires
a deep attention because of the complexity of $\psi_{\sf Jain}^{p,n}(z,\bar{z})$
and therefore needs to be done separately. Also another important point that is
one could think about a relation between these potentials and those
of Chern--Simons in similar way to the Laughlin case.

\section{$SU(N)$ wavefunctions}

Using different theoretical arguments, some authors suggested a
possible FQHE in graphene. As we have seen before, this will be
natural if we look at the anomalous IQHE as a product of
 collective behavior of the composite fermions instead of
particles. Moreover, different wavefunctions have been proposed
to describe the anomalous FQHE for different filling factors. Among them, we cite
that constructed by Goerbig and Regnault~\cite{goerbig2} to solve some
problems brought by other theories like for instance that
developed in~\cite{toke}. However, this wavefunction is
sharing many common features with our early proposal~\cite{schreiber}.
On the other hand, the  Goerbig and Regnault states are a direct extended
version of those of the Laughlin~\cite{laughlin} as well as Halperin~\cite{halperin}.
To get more advantages, let us
return back to the constructed $SU(N)$ wavefunctions~\cite{jellal} describing
the anomalous FQHE at the filling factor 
\begin{equation}
\label{biff}
\nu=q_iK_{ij}^{-1}q_j
\end{equation}
where $K_{ij}$ is an $N\times N$ matrix and $q_i$ is a vector.
In fact, this included different fractions suggested recently
for FQHE in graphene and allowed to make contact
with different proposals.


We start by reviewing the building of such wavefunctions. Indeed,
to construct a general wavefunction, we use the obtained
result so far. 
In doing so,
we consider two sectors labeled by $(m)$ and $(n)$. This is the case for
instance in graphene where there are two subsystems forming a honeycomb.
Le us define $\psi^{(m,n)}$ as a tensor product
\beq
\psi^{(m,n)} = \psi^{(m)} \otimes \psi^{(n)}
\eeq
where each $\psi^{(m)}$ is given in~(\ref{eigfunf}). Assuming that
the condition between matrix elements $K_{ij}=K_{ji}$ is
fulfilled, a natural way to construct the required
 wavefunction is
\begin{eqnarray}
\label{nvacuum}
| \Psi \rangle_{\sf Jel} &=& \prod_{m=1}^{N}
\left[ \epsilon ^{i_1\cdots i_{M_m}} \psi_{i_1}^{(m)} \psi_{i_2}^{(m)}
\cdots \psi_{i_{M_{m}}}^{(m)}
\right]^{K_{mm}-K_{mn}}  \ \prod_{n=1}^{N}
\left[ \epsilon ^{j_1\cdots j_{M_n}} \psi_{i_1}^{(n)} \psi_{i_2}^{(n)}
\cdots \psi_{i_{M_{n}}}^{(n)}
\right]^{K_{nn}-K_{mn}}
\nonumber\\
&&
\prod_{m<n}^{N}
\left[ \epsilon ^{k_1 \cdots k_{M_m+M_n}} \psi_{k_1}^{(m,n)} \psi_{k_2}^{(m,n)}
 \cdots \psi_{i_{M_{m}+M_{n}}}^{(m,n)}
\right]^{K_{mn}}  | 0 \rangle.
\end{eqnarray}
It is not hard to get  the
constraint
\begin{equation}
L_{\sf tot}|\Psi \rangle_{\sf Jel} =0
\end{equation}
where the total angular momenta is given by
\beq
L_{\sf tot}=\sum_{i}^N L_{i}.
\eeq
This wavefunction is actually capturing some interesting
effect that is relevant to discuss about
FQHE in graphene. Indeed, the following term
\begin{equation}
\lb{INT}
\prod_{m<n}^{M}
\left[ \epsilon ^{k_1 \cdots k_{M_m+M_n}} \psi_{k_1}^{(m,n)} \psi_{k_2}^{(m,n)}
 \cdots \psi_{i_{M_{m}+M_{n}}}^{(m,n)}
\right]^{K_{mn}}
\end{equation}
is nothing but is an inter-layer correlation. In conclusion, our
configuration could be a good ansatz for
the ground states of FQHE in graphene. This
will be clarified soon.

At this level, we have all ingredients to do our job. Indeed,
we start by defining
a new complex variable
\begin{equation}
\label{zeta}
\zeta_i=\left\{
\begin{array}{l}
z_{i}^{(m)} \qquad {\mbox{for}}\; i=1, \cdots, M\\
z_{{i-M}}^{(n)}\qquad {\mbox{for}}\; i=M+1,  \cdots,2M
\end{array}
\right.
\end{equation}
assuming that the particle numbers are
equal, i.e. $M_1=M_2=M$,
and recalling the antisymmetric Vandermonde determinant
for the fully occupied state
\begin{equation}
\prod_{i<j} \left(z_i-z_j\right) = {\rm det}\left(z_i^{M-j}\right)=
\epsilon^{i_1 \cdots i_M}z_{i_1}^{0}\cdots z_{i_M}^{M-1}.
\end{equation}
Therefore,~(\ref{nvacuum})
can be projected on the complex plane as
\begin{eqnarray}
\label{projection}
\Psi_{\sf Jel} &=&
\prod_{m=1}^N  \left[ \epsilon ^{i_1\cdots i_M}  \left(z_{i_1}^{(m)}\right)^0 \cdots
 \left(z_{i_M}^{(m)}\right)^{M-1} \right]^{K_{mn}-K_{mn}}
\prod_{n=1}^N  \left[ \epsilon ^{j_1 \cdots j_M}  \left(z_{j_1}^{(n)}\right)^0 \cdots
\left(z_{j_M}^{(n)}\right)^{M-1} \right]^{K_{nn}-K_{mn}}
\nonumber\\
&&
\prod_{m<n}^N \left[ \epsilon ^{k_1 \cdots k_{2M}}  \zeta_{k_1}^0 \cdots
\zeta_{k_{2M}}^{2M-1} \right]^{K_{mn}}\;
\Psi_{0}.
\end{eqnarray}
It can be written in the standard form as
\beq\lb{zcvacu}
\Psi_{\sf Jel}=
\prod_{m=1}^N \prod_{i<j}^M \left(z_i^{(m)} - z_j^{(m)} \right)^{K_{mm}}
 \prod_{n=1}^N \prod_{i<j}^M \left(z_i^{(n)} - z_j^{(n)} \right)^{K_{nn}}
\prod_{m<n}^N \prod_{i,j}^M \left(z_i^{(m)} -z_j^{(n)} \right)^{K_{mn}}
\; \Psi_{0}.
\eeq
This exactly coincides with that constructed by Goerbig and Regnault~\cite{goerbig2}
and what we have proposed~\cite{schreiber} in terms of the matrix model and
non-commutative Chern--Simons theories.
The wavefunction~(\ref{zcvacu}) is a good candidate for describing the anomalous
FQHE in graphene. As it is seen in~\cite{jellal}, the present wavefunction is
general and has many features. In particular, it can be fixed to
 recover different filling factors like for instance
those in~\cite{halperin,girvin1} as well as others.
Note that, the above wavefunctions are the planar limit
of those constructed on two-sphere~\cite{jellal}.

As far as the graphene is concerned, we restrict our analysis to
the case $N=4$. In fact, we give an explicit realization of the Jain potential
and therefore the Hamiltonian that captures such interactions.
In doing so, we can use the same procedure as for the Laughlin case
to write down the potential behind~(\ref{zcvacu}). For instance,
the first term in~(\ref{zcvacu}) is originated from
\begin{equation}\lb{jeloinj}
[(V_z^{\sf Jel})_i]_{|_{K_{mm}}}= -2i{\pa\over \pa z_i} (\ln  F^{\sf Jel}_{K_{mm}} -
\ln\bar F^{\sf Jel}_{K_{mm}}), \qquad
[(V_{\bar z}^{\sf Jel})_i]_{|_{K_{mm}}}= 2i{\pa\over \pa \bar z_i} (\ln  F^{\sf Jel}_{K_{mm}} -
\ln\bar F^{\sf Jel}_{K_{mn}})
\end{equation}
and the potential corresponding to the second term can be
obtained in similar way, i.e. changing $K_{mm}$
by $K_{nn}$. However, the inter-layer interaction is
\begin{equation}\lb{jeloinji}
[(V_z^{\sf Jel})_i]_{|_{K_{mn}}}= -2i{\pa\over \pa z_i} (\ln  F^{\sf Jel}_{K_{mn}} -
\ln\bar F^{\sf Jel}_{K_{mn}}), \qquad
[(V_{\bar z}^{\sf Jel})_i]_{|_{K_{mn}}}= 2i{\pa\over \pa \bar z_i} (\ln  F^{\sf Jel}_{K_{mn}} -
\ln\bar F^{\sf Jel}_{K_{mn}}).
\end{equation}
These can be worked as for the previous cases
to get the corresponding wavefunctions. In doing so, we
end up with the form
\begin{equation}\lb{jelwf}
\psi_{\sf Jel}^{K_{mm},K_{mn}}(z,\bar{z})=
\prod_{m=1}^4 F^{\sf Jel}_{K_{mm}}  F^{\sf Jel}_{K_{nn}} F^{\sf Jel}_{K_{mn}}
\; \Psi_{0}.
\end{equation}
Of course this is associated to a  Hamiltonian,
which involves such potentials. Indeed, this can be
obtained as two copies of $H_{\sf L}^{\sf int}$ amended by inter-layer
correlation. It can straightforwardly be generalized to include
other kind of interaction, in particular more that
three-body type.

The above wavefunctions has many features. Indeed, it can be linked to the
composite Dirac fermions
and derive some interesting results. To clarify this point, we return to the
definition of the filling factor and write for any $j$ the relation
\beq\lb{jff}
\nu_j = {\rho_j\over N_B}
\eeq
where the number $N_{B}$ is the quantized flux. In our case, it can
be written in terms of elements
of the matrix $K$ (\ref{biff})
as
\begin{equation}\label{qufl}
N_B = \rho_j K_{j} +\sum_{i\neq j} \rho_i K_{ij}
\end{equation}
and we have set $K_{jj}=K_{j}$. This offers different discussions
about the above wavefunctions and its links to other proposals. For instance, we can
establish a bridge between the filling factor (\ref{biff}) and that
coresponds to the composite Dirac fermions.
For this, let us start by identifying (\ref{nunus}) and (\ref{biff})
to see what are the consequences. Clearly, for $n=0$
we end up with the Laughlin series, namely
\beq\label{lser}
{1\over K_{ij}} = {2\over 4p \pm 1}
\eeq
which is in agreement with different investigations~\cite{toke2,jellal}.
However, the interesting case is when $p=0$ that leads to the
relation
\begin{equation}\label{mapff}
2\sum_{i\neq j} {\rho_i\over \rho_j} K_{ij}= \pm\left({1\over 2n+1} -1\right).
\end{equation}
This offers different discussions about the ratio ${\rho_i\over \rho_j}$
and in particular its quantization. For this,
let us explicitly evaluate the involved series,
which can be done by recalling a similar
formula. This is~\cite{grad}
\begin{equation}\label{forff}
\sum_{l=1}^{m} {(-1)^{k+1}\over k+1}{m!\over k!(m-k)!} ={1\over m+1}-1
\end{equation}
 Therefore,
by comparing both series we can identify $2n$ to $m$ and
more importantly the ration
can be quantized as follows
\beq
{\rho_i\over \rho_j}\sim {(-1)^{k+1}\over 2(k+1)}{2n!\over k!(2n-k)!}K_{ij}^{-1}.
\eeq
This mapping shows how the present theory can be restricted to make
a possible connection with the composite Dirac fermions.
Moreover, other theories can be recovered from our the
present results like for instance
that proposed in the reference~\cite{toke2}.
On the other hand, some discussions
can be added about the nature of the Hall
droplet in terms of the ratio and in particular
the quantization of its area.

\section{Conclusion}

The present work was devoted to analyze the behaviour of an interacting
system in the presence of an external magnetic field. More precisely,
we have focused on a specific case where the system is exactly a sheet of graphene.
To do this, we have started by considering a model that captures a kind of
interaction, which applied before at different occasions to
discuss the conventional QHE. Before doing so, we have presented some discussions
about the effective potential that corresponds to the Laughlin
wavefunction at the filling factor $\nu={1\over 2l+1}$. This was
a typical example of a generic case that is the Jain
wavefunction.


By considering a Hamiltonian of the present system where the interaction
is restricted to be two and three-body types, we have employed a previous
technical to achieve our goal. This was done by making use the mapping between
three Hamiltonian's: Landau, Pauli--Schr\"odinger and Dirac. This means that
we have derived an appropriate Dirac operator in terms of the involved interaction.
The mapping allowed us to
get in a simple way different spectrum's entering in the game and the excited states
as well. Using a gauge transformation,  we have established a link between the free
and interacting Dirac operators. This can be generalized to take into
account of other interaction types.

As concerning QHE in graphene, in the beginning we have extended the Laughlin potential
to the Jain case. Indeed, we have generalized the interaction terms that are behind the
series $\nu={1\over 2l+1}$ to that for the composite fermions in graphene, i.e.
$\nu^{\sf G}$. This was shown another way to talk about the Jain wavefunction for
graphene and its corresponding effective potential. Of course these can be used to
deal with other proposals related to different wavefunctions suggested for
the anomalous FQHE.

As second part, we have concentrated on the $SU(N)$ wavefunction
that captures different filling factors and generalizes some
others. In the first step, we have given the corresponding potentials
as an explicit example to illustrate our general procedure for
Jain wavefunction.
To explicitly show its relation to some theories, we have
established its link to the composite fermion pictures. This has been
done by identifying both of filling factors. As consequence, we have found
a quantized term that include the density of interacting particles. This
in principle gave rise to another way to interpret our results and made
contact with others.

The obtained results suggest to investigate and
deal with other issues related to QHE in graphene. As example,
we are now focussing on a second problem that involves another
type of interaction. More precisely, we are wondering to understand
the behaviour of the confined relativistic particles in the
presence of an external magnetic field.

\section*{Acknowledgment}

This work was completed during a visit of AJ to the Abdus Salam Centre for Theoretical
Physics (Trieste, Italy) in the framework of junior associate scheme.
He would like to acknowledge the
financial support of the centre.

\end{document}